\documentstyle[aps,prl,preprint]{revtex}
\begin{document}
\draft
\title{First order wetting of rough substrates and quantum unbinding}
\author{Attilio L. Stella}
\address{INFM-Dipartimento di Fisica and Sezione INFN, Universit\`a 
di Padova, I-35131 Padova, Italy}
\author{Giovanni Sartoni}
\address{Instituut--Lorentz, Rijks Universiteit Leiden, 2300 RA Leiden,
The Netherlands}
\date{\today}
\maketitle
\begin{abstract}
Replica and functional renormalization group methods show
that, with short range substrate forces or in strong fluctuation regimes, 
wetting of a self--affine rough wall in $2D$ turns first--order
as soon as the wall roughness exponent exceeds the anisotropy index of
bulk interface fluctuations. Different thresholds apply with long
range forces in mean field regimes. For bond--disordered bulk, fixed point
stability suggests similar results, which ultimately
rely on basic properties of quantum bound states with asymptotically
power--law repulsive potentials. 
\end{abstract}
 

Wetting transitions occur when, e.g., an interface separating two coexisting 
phases unbinds from an attractive substrate. In the recent literature,
much space has been devoted to the effects of
different types of disorder on the nature and universality of such
transitions \cite{forgacs91,fisher86,kardar85,fisher87}. 
This is motivated both by the presence of impurities in
actual experiments, and by the expectation that disorder modifies critical
behavior. Many studies concentrated on impurities in the bulk, or on the 
surface of a smooth substrate. Another type of disorder is that due to
the roughness of the wall delimiting the substrate. This rather frequent
geometrical disorder was discussed 
especially in connection with measurements of nitrogen adsorption on
flash deposited silver\cite{pfeifer89,okki90,li90,stella96}.

Substrate roughness describable by self--affine geometry is often realized
and most interesting, from both a fundamental, and a physical point of view.
Indeed, a wall whose average transverse fluctuation, $W_L$, increases as a
power of the longitudinal sample size, $L$ ($W_L\propto L^{\zeta_w},
0<\zeta_w<1$), has a random geometry characterized globally by a single
roughness exponent, $\zeta_w$. Moreover, no matter whether their fluctuations
are controlled by temperature or by disorder, bulk interfaces behave
as self--affine objects, with appropriate exponents $\zeta_0$
\cite{forgacs91,fisher87}.
Thus, one may expect a direct competition between wall and interface 
roughnesses to take place in wetting phenomena, when substrates of this kind 
are considered.
 
In the present Letter we show exactly in $2D$ that, as soon as the wall
wins (i.e. for $\zeta_w>\zeta_0$ with short range substrate potentials), 
the above competition
is resolved in an unusual, drastic change of the wetting transition, from
continuous to first--order. We determine exact roughness thresholds 
for first--order wetting also in other regimes with the substrate
exerting long range forces (e.g., van der Waals) on the interface
and discuss, at perturbative level, cases with bond--disorder in the bulk.
A change from continuous to discontinuous wetting is in fact
a quite surprising and unexpected phenomenon, especially in $2D$. 
As a rule, with
short--range forces, first--order wetting never occurs in $2D$, except in
special and ad hoc limit situations\cite{nieuwen88}.
Disorder generally acts in the opposite sense of turning into second--order
discontinuous transitions\cite{forgacs91}. 
In all cases the mechanism leading to
first--order wetting at large $\zeta_w$ can be traced back to some
basic properties of quantum bound states of a particle in a potential
\cite{zia88}. Such properties turn out to have
very far reaching consequences for our understanding of
wetting phenomena.

In $2D$, a self--affine wall can be described by a random function
$h_w(x)$ (Fig. 1), with probability distribution\cite{li90}
\begin{equation}\label{1}
P_w[h_w] \propto \exp{[-\int dx {K_w\over 2} ({\partial}^{\beta}h_w/\partial x
^{\beta})^2 }]
\end{equation}
such that $\overline{|h_w(x)-h_w(x')|}\propto |x-x'|^{\zeta_w}$, with
$\zeta_w=\beta-1/2$ and the overbar indicating quenched average with
respect to (1)\cite{nota0}. In the presence of
bulk disorder, the interface Hamiltonian can be put in the form
\begin{eqnarray}\label{2}
\beta H[h,h_w,V]&=& \int dx [{K \over 2} (\partial h/\partial x)^2+ 
U(h(x)-h_w(x))+\nonumber\\
&&+V(x,h(x))]
\end{eqnarray}
where $V$ is a Gaussian random potential ($\overline{V}= 0$, $\overline
{V(x,h)V(x',h')}=\Delta \delta(x-x') \delta(h-h')$), and $U$ is the
potential due to the substrate. $K$ is the interfacial
stiffness. If the wall is attractive, but
impenetrable, $U(y)=\infty$ for $y\leq 0$, and a minimum of $U$ at some
$y>0$ allows to pin the interface. With long range forces, 
$U(y)\sim u/y^{\sigma-1}+v/y^{\sigma}$ ($v>0$), for large $y$\cite{fisher87}.

The disorder due to $h_w$ and $V$ in eq.(2) makes the partition 
$Z[h_w,V]=\int {\cal D}h \exp(-\beta H)$ a stochastic
variable. Thus, we introduce replicas\cite{replica} and evaluate 
\begin{equation}\label{3}
\overline{Z^n}=\int {\cal D}V \int {\cal D}h_w P_v[V] P_w[h_w] 
\int \prod_{\alpha=1}^n {\cal D}h_{\alpha} \exp(-\beta H[h_{\alpha},h_w,V])
\end{equation}
where $\ln(P_v)=const-{ 1\over 2\Delta} \int dx dh V(x,h)^2$. Integration over
${\cal D}V$ is easily performed and allows to interpret 
$h_{\alpha}(x)$ as world lines ($x$ corresponding to time) of $n$ quantum
particles interacting via attractive two--body $\delta$--potentials. Thus, in
eq.(3) we are left with integrations over ${\cal D}h_w$ and ${\cal D}h_\alpha$,
and an effective Hamiltonian:
\begin{eqnarray}\label{4}
\beta H[h_{\alpha},h_w]&=& 
\int dx \left( {K_w \over 2} ({\partial}^{\beta} {h_w}/{\partial x}^{\beta})^2+
\sum_{\alpha}[{K\over2}(\partial h_{\alpha}/\partial x)^2+
{K'\over2}(\partial h_w/\partial x)^2+\right.\\
& &\left.C(\partial h_{\alpha}/\partial x)
(\partial h_w/\partial x)+U(h_{\alpha})]+
\Delta \sum_{\alpha\neq\beta}\delta(h_{\alpha}-h_{\beta}) \right)\nonumber
\end{eqnarray}
The logarithm of $P_w$ is now included into $\beta H$ and 
the couplings $K'$ and
$C$ arise from the replacement $h_{\alpha} \to h_{\alpha}+
h_w$ (initially, of course, $K'=C=K$).
A functional renormalization group (RG)\cite{forgacs91,li90} 
treatment can be performed exactly
up to first order in $U$ and $\Delta$. By summing up $\exp(-\beta H)$ over
Fourier modes $\tilde{h}_{\alpha}(k)$ and $\tilde{h}_w(k)$ 
with $\Lambda/b<k<\Lambda$,
after the rescalings $x \to bx$, $h_{\alpha}\to b^{\zeta}h_{\alpha}$ and
$h_w \to b^{\zeta_w}h_w$, one obtains the following RG flow equations
( $b=1+dl$):
\begin{eqnarray}\label{5}
{d\ln(U) \over dl}&=&1+\zeta h{U' \over U}+\Omega{U'' \over U}\\
{d\ln(K) \over dl}&=&2 \zeta-1\nonumber\\
{d\ln(K') \over dl}&=&2\zeta_w-1\nonumber\\
{d\ln(C) \over dl}&=&\zeta+\zeta_w-1\nonumber\\
{d\ln(K_w) \over dl}&=&1-2\beta+2\zeta_w=0\nonumber\\
{d\ln(\Delta) \over dl}&=&1-\zeta\nonumber
\end{eqnarray}
where $\Omega$ is a suitable function of $K$, $K_w$, $C$ 
and the cut--off $\Lambda$\cite{li90}, and 
$h$ stands for a generic $h_{\alpha}$.

We first discuss $\Delta=0$. With
$\zeta_w<1/2$ and ordered bulk, for $\zeta=\zeta_0=1/2$
\cite{forgacs91,fisher87},
there exists a fixed point (FP) 
of eqs.(5),
with respect to which $K'$ and $C$ are irrelevant ($d\ln(K)/dl=0$, while,
e.g., $d\ln(K')/dl<0$). Thus, substrate fluctuations
decouple from the problem.
With short range, or with long range forces such that $\zeta_0=1/2
>2/(\sigma+1)=\zeta^*$ (strong fluctuation (SF) regime\cite{fisher87}),
the FP behavior of $U$ at large $h$ is within the control of
a first order cumulant expansion and turns out to 
be $U \propto h^{-\tau(\zeta_0)}$, 
($\tau(\zeta_0)=2(1-\zeta_0)/{\zeta_0}=2$)\cite{fisher87,li90,nota1}.
This long distance behavior should in fact apply to
all the FP $U$'s necessary to describe the wetting transition 
in such conditions.
These FP's are in general three: one describing pinned interface
situations, one for the wet regime with unbound interface, and one, unstable,
at the borderline between the domains of attraction of the previous two,
describing the transition point behavior.
In view of the decoupling of substrate fluctuations,
the wetting transition controlled by these FP's, 
whose $U$'s we can not determine at finite $h$, 
is expected to be continuous, with exponents identical to those valid for flat 
wall, which are known exactly \cite{forgacs91,abrahams}.
In the case of short range forces, numerical evidence of second order wetting
with such exponents has been recently obtained for low enough
$\zeta_w$ by extensive transfer matrix calculations\cite{stella96}.

The FP's for $\zeta_w>1/2$ have to be found at $T=0$, by
putting $\zeta=\zeta_w$ in eqs.(5). 
Indeed, now, choosing again $\zeta=1/2$, 
parameters like $K'$ and $C$ would grow to infinity while $K$
remains fixed. Surface roughness is clearly relevant now. Under a rescaling 
$b$, a $T=0$ fixed point is approached as $\beta H \propto b^y (\beta H)^*$
with $(\beta H)^*$ finite and $y>0$, when $b\to \infty$. Such FP's
are expected in situations 
when quenched disorder (due to the wall here) controls the physics\cite{li90}.
At the $T=0$ FP's with $\zeta=\zeta_w$, $K$, $K'$ and $C$ are all
growing to infinity at the same rate ( e.g., $K(l) \sim K^* 
\exp[(2\zeta_w-1)l]$), and $U(l) \sim U^*(h) \exp[(2\zeta_w-1)l]$.
$U^*$ obeys an equation like the first of eqs. (5), with the constant 
term replaced
by $2(1-\zeta_w)$, and $\zeta_w$ multiplying the second term on the r.h.s.
in place of $\zeta$. Thus, the discussion of the asymptotic behavior of $U^*$
follows lines similar to those for $U$ in the case $\zeta_w<1/2$\cite{li90}. 
In particular, with short range forces or in SF regime, 
we get now $U^*(h)\propto h^{-\tau(\zeta_w)}$, ($\tau(\zeta_w)<2$). Such
behavior of $U^*$ holds also
in MF regime ($\zeta^*>\zeta_0=1/2$\cite{fisher87}), as soon 
as $\zeta_w>\zeta^*$. 

This asymptotic behavior of $U^*$ and the connection between path integral
and quantum mechanics are the key to demonstrate
first--order wetting. Indeed, the transition order
is revealed by the way in which $\overline{\langle h \rangle}$ 
diverges to infinity. 
Consistently with eqs. (5), upon approaching a $T=0$ FP with 
$\zeta_w>\zeta_0$, or $\zeta_w>\zeta^*$ in MF regime, we must define 
${K_w}^*$ such that $K_w(l)={K_w}^*\exp[2\zeta_w-1)l]=const$.
Thus, in the FP action $(\beta H)^*$
we are left with $K_w^*=0$, as $l \to \infty$. By shifting back
integration variables in this action ($h_{\alpha} \to h_{\alpha}-h_w$), the
terms in $K'^*$ and $C^*$ disappear and the calculation of $\overline{Z^n}$
can be easily converted into that of the ground state energy
of a quantum problem in $1D$, with $n+1$ particles and Hamiltonian
${\cal H}= \sum_{\alpha}[p_{\alpha}^2/(2{K^*})+U(h_{\alpha}-h_w)]$. 
In this problem the 
particle with coordinate $h_w$ has an infinite mass. This
circumstance allows to get the ground state wave function of ${\cal H}$
exactly in the form $\prod_{\alpha} \Psi(h_{\alpha}-h_w)$, 
with $\Psi$ satisfying the one--particle Schr\"odinger equation:
\begin{equation}\label{6}
-{1 \over {2K^*}} \partial^2 \Psi/{\partial h^2}+U^*\Psi=\epsilon \Psi .
\end{equation}
$\overline{\langle h_{\alpha}-h_w\rangle}$ is proportional 
to the expectation value, 
$\langle h \rangle_{\Psi}$\cite{nota2}, of $h$ in the ground--state $\Psi(h)$
of eq.(\ref{6}).
We concluded above that, at large $h$ and for $\zeta_w>1/2$ 
(or $\zeta_w>\zeta^*>1/2$
with long range forces in MF regime), the possible
$U^*(h)$, however behaving at finite $h$, are repulsive and
decay asymptotically to zero with a power $\tau(\zeta_w)<2$ of $h$. 
The FP $U^*$ at the wetting transition must
have such a shape to belong to the class of borderline between
potentials with bound ground state and $\epsilon<0$, and potentials 
for which all states have $\epsilon>0$ and $\langle h \rangle_{\Psi}=\infty$.
These two latter types of potentials characterize 
dry and wet regimes, respectively.
Independent of the details of $U^*(h)$ at short $h$, a solution 
of eq.(\ref{6}) with
$\epsilon=0$ has a remarkable property for $\tau<2$\cite{zia88}. 
Indeed, an $\epsilon=0$ eigenstate necessarily behaves as
$\Psi(h) \propto \exp(-a h^s)$, at large $h$, with $s=1-{\tau \over 2}>0$.
This means that, for $\epsilon=0$, a repulsive
potential decaying slower than $h^{-2}$ creates a too strong barrier at
large distances, to allow interface delocalization. Thus, the ground 
state $\Psi$ for $U^*$ representing the transition FP (i.e. a FP
potential in the borderline class) must be bound, with ${\langle h
\rangle}_{\Psi}<\infty$. This implies that, right at the wetting
transition, $\overline{\langle h \rangle}<\infty$, while
$\overline{\langle h \rangle}=\infty$ as soon 
as the wet phase is accessed.
First--order wetting is thus proved as soon as $\zeta_w>1/2$
( short range forces or $\zeta^*<1/2$), or $\zeta_w>\zeta^*>1/2$. 

A recent numerical study of a $2D$ model with rough
substrate exerting short range forces, gave evidence in support of
first--order wetting for $\zeta_w$ sufficiently larger than $1/2$
\cite{stella96}. In order to get a more direct manifestation of the
mechanisms implied by eq.(6), we performed transfer matrix calculations for
a model on square lattice with both wall and interface represented by 
directed paths, as described in ref.\cite{preprint}.
Fig.1 reports numerical results for the probability distribution
of $h$. Data are taken just below
the depinning temperature for $\zeta_w=2/3$ ($\tau(2/3)=1$). The dotted curve
has a behavior $\propto \exp(-a x^{1/2})$, of the form expected
right at threshold on an infinite asymptotic range ($s=1/2$). 
A relatively still poor sampling over disorder is largely responsible of
some oscillations of the distribution, but the overall
trend appears already consistent with our theoretical predictions.

With bulk disorder ($\Delta>0$), the perturbative character
of eqs.(5) prevents an exact control of the FP $U$
for $h \to \infty$. On the other hand, we know that, with
$\Delta>0$, $\zeta_0=2/3$ is the exact interface anisotropy 
index\cite{forgacs91,fisher87}. 
For $\zeta_w<2/3$, by putting $\zeta=2/3$ in eqs.(5),
we find that both $K(l)$ and $\Delta(l)$ grow proportional to $\exp(l/3)$
(towards a $T=0$ fixed point), while $K'$ and $C$ grow slower, and are
thus irrelevant. At the same time,
for short range forces, $U(l)=U^*\exp(l/3)$ gives $U^*(h)\propto
h^{-\tau(\zeta_0)}$, with $\tau(\zeta_0)=1$, at large $h$. Thus, in the 
limit of very small
bulk disorder, we get indication that for $\zeta_w<2/3$ a wetting transition
regime identified by $\zeta=\zeta_0=2/3$ should imply a decoupling
of substrate fluctuations from the problem.
At least with very weak bulk disorder, the wetting transition
with $\zeta_w<2/3$ should retain the features of the flat wall case.
For this case one indeed expects an effective wall--interface potential 
decaying
as $h^{-1}$\cite{fisher86,fisher87}, and Kardar has determined 
exactly by Bethe
ansatz the second order character and the exponents of the transition
\cite{kardar85}. Consistently with our expectation,
numerical results for short range forces in ref.\cite{preprint}
support continuous wetting in Kardar's class
for $\zeta_w$ sufficiently lower than $2/3$, 
even with finite bulk disorder.

Let us consider now $\zeta_w>2/3$, and short range forces again. 
By putting $\zeta=\zeta_w$ in eqs. (5), we
find $\Delta(l)=\Delta(0)\exp[(1-\zeta_w)l]$, while $K(l)=K'(l)=C(l)=
K^* \exp[(2\zeta_w-1)l]$ and $K_w(l)=const$. Furthermore, 
$U(l)=U^*\exp[(2\zeta_w-1)l]$ implies $U^*(h)\propto h^{-\tau(\zeta_w)}$.
Since now $\Delta$ (still supposed small) grows slower than $K, C$ and $K'$,
it is natural to regard it as an irrelevant parameter with respect to 
the $T=0$ FP's which would be reached for $\Delta=0$ strictly.
Upon varying $\zeta_w>2/3$, these FP's span a subset of those
already discussed with ordered
bulk, for which quantum mechanics implies first--order wetting. Thus, we
conclude that for $\zeta_w>2/3$ a small amount of bulk disorder is
irrelevant and leaves the transition under control of the same mechanism
outlined for pure bulk and the same $\zeta_w$. Numerical results
in ref.\cite{preprint} support this
conclusion, giving evidence of first--order wetting
for sufficiently large $\zeta_w$ and finite disorder. Similar arguments
apply to long range forces in SF and, for $\zeta_w>\zeta^*$,
in MF regime. 

In summary, our RG picture demonstrates first--order wetting in $2D$
with sufficiently 
rough substrates exerting short range forces on the interface. 
This is consistent with earlier
numerical work suggestive of discontinuous transitions
\cite{stella96,preprint}. 
The threshold for first--order wetting is precisely identified 
as $\zeta_w=1/2$ in the case of
ordered bulk. For disordered bulk perturbative arguments
suggest first--order as soon as $\zeta_w>2/3$, consistent with
a possible general rule that $\zeta_0$ identifies the threshold.
We predict roughness induced first--order wetting
also with long range forces,
for $\zeta_w>\zeta_0>\zeta^*$ (SF) or for $\zeta_w>\zeta^*>\zeta_0$ 
(MF).
Discontinuous depinning is due
to the repulsive effective wall--interface potential, which 
becomes too strong, at large distance, to allow for a continuous 
increase towards
infinity of $\overline{\langle h \rangle}$ when depinning
is approached. This follows from general quantum properties,
independent of the details of U at finite $h$.

Interesting open problems remain the nature of wetting right at the thresholds
and the possible extension to $3D$ of this type of results, which rely
on the connection with quantum mechanics in $1D$.
A recent mean feld study in $3D$ suggests the possibility of
first--order wetting induced by wall roughness with short range substrate
potential and ordered bulk\cite{parry96}. Another interesting issue is whether
$\zeta_w=\zeta_0$ could be a plausible threshold 
also in cases in which different kinds of bulk disorder imply 
different $\zeta_0$'s.
Relevant examples include random fields \cite{fisher87} and 
quasicristals \cite{lipowsky91}.

Due to the competition between two qualitatively similar scaling geometries,
interactions between a fluctuating 
manifold and a random boundary can lead to interesting phenomena
also in other contexts. An example could be
flux lines in high--$T_c$ superconductors with extended rough 
defects\cite{deloc}. 
Also of interest would be polymers or membranes adsorbed by
rough walls. 

We thank M. Kardar for useful criticism and G. Giugliarelli for ongoing
collaboration. G.S. was supported by the Stichting voor Fondamenteel
Onderzoek der Materie.

\begin{figure}
\caption{ Inset: sketch of the geometry of wall (continuous) and interface (dashed).
Main: probability distribution of $h$, from a sampling of
500 wall configurations of $100000$ longitudinal steps.}
\end{figure}


\begin{references}
\bibitem{nota0}Of course there are choices other than (1) giving
the same $2$--point height correlation. Our Gaussian form is the most simple,
with all higher cumulants equal to $0$.
\bibitem{nota1}For $\zeta_0<\zeta^*$, a mean field (MF) regime is realized and 
$U \propto h^{-\sigma+1}$, with an amplitude growing under renormalization.
\bibitem{nota2}In eq.(7) and from here on we use again $h$ to indicate
$h_{\alpha}-h_w$.
\end{references}
\end{document}